\documentclass[11pt,preprint]{aastex}

\begin{document}

\title{ Hypervelocity Stars and the Galactic Center }

\author{Warren R.\ Brown}

\affil{Smithsonian Astrophysical Observatory, 60 Garden St, Cambridge, MA 02138}
\email{wbrown@cfa.harvard.edu}

\section{INTRODUCTION}


	A massive black hole (MBH) will inevitably unbind stars from the Galaxy.  
\citet{hills88} coined the term ``hypervelocity star'' (hereafter HVS)  to describe
a star ejected at $\sim$1000 km s$^{-1}$ from a three-body interaction with a MBH.  
HVSs are a general consequence of MBHs in any galaxy \citep[e.g.][]{sherwin08}, but
the HVSs we can observe are the stars ejected from our own Galaxy.

	\citet{yu03} predict that Sgr A$^*$ ejects one HVS every $\sim$$10^5$ yrs.  
Thus HVSs are rare:  of the Galaxy's $10^{11}$ stars, there are only $\sim$$10^3$
HVSs within 100 kpc of the Milky Way.  Yet HVSs would be very interesting to find
because HVSs provide unique constraints on the nature and environment of the central
MBH.

	In 2005 we reported the discovery of the first HVS:  a 3 M$_{\sun}$ main
sequence star traveling with a Galactic rest frame velocity of at least $+709\pm12$
km s$^{-1}$, more than twice the Milky Way's escape velocity at the star's distance
of 110 kpc \citep{brown05}.  This star cannot be explained by normal stellar
interactions: the maximum ejection velocity from binary disruption mechanisms
\citep{blaauw61, poveda67} is limited to $\sim$300 km s$^{-1}$ for 3 M$_{\sun}$
stars \citep{leonard88, leonard90, leonard91, leonard93, tauris98, portegies00,
davies02, gualandris05}.  Thus a massive and compact object is needed to accelerate
a 3 M$_{\sun}$ star to an unbound velocity.

	There is overwhelming evidence for a $\sim4\times10^6$ M$_{\sun}$ MBH in the
dense stellar environment of the Galactic center \citep[e.g.][]{schodel03, ghez05}.   
	The three-body exchange mechanism that may explain S-stars orbiting the MBH
\citep{gould03b} is the same mechanism that may eject HVSs \citep{hills88}.  In this
picture, S-stars are the former companions of HVSs \citep{ginsburg06, ginsburg07}.
	Interestingly, the S-stars are main sequence B stars \citep{ghez03,
eisenhauer05, martins08}, just like the observed HVSs described below.

	In \S 2 we define HVSs in the context of unbound stars, and discuss how HVSs
are made.  In \S 3 we highlight recent HVS discoveries and discuss their nature.  
In \S 4 we explore the links between HVSs and the Galactic Center.  We conclude in
\S 5.

\section{THEORETICAL CONSIDERATIONS}

\subsection{Defining an Hypervelocity Star}

	Following \citet{hills88}, we define HVSs by 1) their MBH origin and 2)
their unbound velocities.  An HVS ejected from the Milky Way travels on a nearly
radial trajectory; the expected proper motion for a 50 kpc distant HVS is a few
tenths of a milliarcsecond per year \citep{gnedin05}.  Thus radial velocity directly
measures most of an HVS's space motion.  Deciding whether an HVS is unbound,
however, requires knowledge of the star's location in the Galaxy.

	Unfortunately, the Galactic potential is poorly constrained at distances
$>50$ kpc.  \citet{kenyon08} discuss a form of the potential that, for the first
time, fits the Milky Way mass distribution from 5 pc to 10$^5$ pc.  Because the
potential does not yield a formal escape velocity, they define unbound stars as
having $v_{rf}>200$ km s$^{-1}$ at $R=150$ kpc.  This conservative definition yields
a Galactic escape velocity of 360 km s$^{-1}$ at 50 kpc and 260 km s$^{-1}$ at 100
kpc.  \citet{xue08}, on the other hand, fit a halo potential model to the velocity
dispersion of 2466 BHB stars located 5 kpc $<R<60$ kpc.  The escape velocity
resulting from the \citet{xue08} model is 290 km s$^{-1}$ at 50 kpc and 190 km
s$^{-1}$ at 100 kpc.

\subsection{Hyper-Runaways}

	Not all unbound stars are necessarily HVSs.  Fast-moving pulsars, for
example, are explained by supernova kicks \citep[e.g.][]{arzoumanian02}.  The star
HD 271791 is the first example of an unbound ``hyper-runaway'' that was ejected from
the outer disk, in the direction of Galactic rotation, when its former 55 M$_{\sun}$
binary companion exploded as a supernova \citep{heber08, przybilla08c}.  Objects
ejected in this manner are traditionally called runaways \citep{blaauw61}.  The term
runaway also includes stars dynamically ejected from binary-binary encounters
\citep{poveda67}.

	Runaway star ejection mechanisms share a common velocity constraint:  the
physical properties of binary stars.  Theoretically, the maximum ejection velocity
from disrupting a binary is the binary orbital velocity, a velocity that increases
with the mass of the stars. The theoretical maximum velocity is not realizable,
however, because compact binaries that are too tight will merge due to energy loss
from tidal dissipation and Roche-lobe overflow \citep[e.g.][]{vanbeveren98}.  While
it is possible for an hyper-runaway to be confused with an HVS in the absence of
proper motions, we estimate that $\sim$3 M$_{\odot}$ HVSs ejected from the Galactic
Center are $\sim$100 times more common than hyper-runaways of the same mass
\citep{brown08c}.  Hyper-runaways are limited by the rarity of massive stars and the
requirement to avoid merging the compact binary progenitor.

\subsection{How to Make Hypervelocity Stars}

	HVSs attain extreme velocities because the gravitational potential energy of
a MBH greatly exceeds the binding energy of a stellar binary.  In Hill's mechanism,
the gravitational tidal force of a single MBH disrupts an approaching binary. One
star is captured on an eccentric orbit around the MBH and, by conservation of
energy, the other star escapes with a final velocity equal to the geometric mean of
the $\sim$$10^4$ km s$^{-1}$ infall velocity \citep[S0-16 had a pericenter velocity
of 12000 km s$^{-1}$,][]{ghez05} and the $\sim$$10^2$ km s$^{-1}$ binary orbital
velocity.  A star traveling $10^3$ km s$^{-1}$ at 1 pc will exit the Galaxy at 100
kpc at 400-500 km s$^{-1}$ \citep{kenyon08}.

	\citet{yu03} further develop Hill's analysis to include the case of a
binary MBH.  While an equal-mass binary MBH is ruled out in the Galactic Center
\citep{reid04}, theorists speculate that the massive star clusters in the Galactic
Center may form intermediate mass black holes (IMBHs) in their cores.  If such IMBHs
exist, dynamical friction will cause them to in-spiral into the central MBH and
eject HVSs along the way.

	The properties of HVSs allow us to discriminate between the single MBH and
binary MBH ejection mechanisms:
	\begin{itemize}
	\item {\it Velocity Distribution.} HVS ejection velocity depends weakly on
the mass of the stellar binary $\propto (m_1+m_2)^{1/3}$ for a single MBH
\citep{hills88}, but there is no such dependence on stellar mass for a binary MBH
\citep[e.g.][]{sesana07b}.
	\item {\it Ejection Rate.} A binary MBH has a larger cross-section and may
eject $\sim$10$\times$ more HVSs than a single MBH \citep{yu03}.
	\item {\it Spatial Distribution.} A binary MBH preferentially ejects HVSs in
its orbital plane, and thus produces a ring of HVSs around the sky
\citep{gualandris05, levin06, sesana06, merritt06}.
	\item {\it Temporal Distribution.} As a binary MBH hardens and then merges,
it will produce a distinctive burst of HVSs over $\sim$$10^7$ yrs, during which HVS
velocities will become systematically more energetic with time \citep{baumgardt06,
sesana06, sesana07}.
	\item {\it Stellar Rotation Distribution.} HVSs ejected by a single MBH
should be slow rotators, because stars in compact binaries have systematically lower
$v\sin{i}$ due to tidal synchronization \citep{hansen07}.  On the other hand, single
stars spun up and ejected by a binary MBH should be fast rotators \citep{lockmann08}.
	\end{itemize}

	HVSs may also be ejected by three-body interactions of single stars
with stellar mass black holes clustered around the central MBH \citep{oleary08}.  
This mechanism predicts that the lowest-mass HVSs will have the highest velocities,
in contrast to Hill's mechanism.

\subsection{Dark Matter}

	One interesting theoretical application for HVSs is as probes of the
Galactic dark matter potential \citep{gnedin05, yu07, wu08}.
	The dark matter paradigm makes specific predictions about the anisotropy of
dark matter halos.  Any deviation of an HVS's trajectory from the Galactic Center
measures this anisotropy.  Unlike tidal streams, HVSs integrate the Galactic
potential out to very large distances.  Interestingly, the \citet{kenyon08}
potential suggests that HVSs may be much more sensitive to the bulge than to the
halo.  Moreover, the bulge potential acts as a high-pass filter: a star must be
ejected at $\sim$800 km/s to reach 1 kpc.

\section{OBSERVATIONS OF HYPERVELOCITY STARS}

\subsection{New HVSs}

	Observers have identified a remarkable number of unbound HVSs in the past 3
years.  Following the discovery of the first HVS \citep{brown05}, \citet{hirsch05}
reported a helium-rich subluminous O star leaving the Galaxy with a rest-frame
velocity of at least $+717$ km s$^{-1}$.  \citet{edelmann05} reported an 9
M$_{\sun}$ main sequence B star with a Galactic rest frame velocity of at least
$+548$ km s$^{-1}$, possibly ejected from the Large Magellanic Cloud (LMC).

	Brown et al.\ designed a targeted HVS survey that has discovered 13-17 new
HVSs, plus evidence for a similar number of bound HVSs ejected by the same mechanism
\citep{brown06, brown06b, brown07a, brown07b, brown08c}.  Briefly, our survey uses
Sloan photometry to select HVS candidates with the colors of late B-type stars
(Figure \ref{fig:spectrum}).  B-type stars have lifetimes consistent with travel
times from the Galactic Center but are not a normally expected Galactic halo
population.  We highlight the latest results of our survey below.


\subsection{Velocity Distribution and Bound HVSs}

	Figure \ref{fig:hist} plots the observed distribution of line-of-sight
velocities, corrected to the Galactic rest-frame \citep[see][]{brown06b}, for the
B-type stars in our survey.  The survey covers 7300 deg$^2$ of sky, with a surface
density of $\sim$0.1 deg$^{-2}$.  The 731 survey stars with $|v_{rf}|<275$ km
s$^{-1}$ have a $-1\pm4$ km s$^{-1}$ mean and a $106\pm5$ km s$^{-1}$ dispersion,
consistent with a normal stellar halo population.

	Remarkably, we observe 26 stars with $v_{rf}>275$ km s$^{-1}$ and only 2
stars with $v_{rf}<-275$ km s$^{-1}$.  The escape velocity of the Milky Way at 50
kpc is $\sim350$ km s$^{-1}$, thus the 12 stars with $v_{rf}>400$ km s$^{-1}$ are
clearly unbound (see also Figure \ref{fig:travel}).
	Ignoring the 12 unbound stars, there is less than a $10^{-5}$ probability of
randomly drawing 14 stars with $275<v_{rf}<400$ km s$^{-1}$ from the tail
of a Gaussian distribution with the observed parameters.  Thus the excess of
positive velocity outliers $275<v_{rf}<400$ km s$^{-1}$ appears significant at the
4-$\sigma$ level.

	The positive velocity outliers demonstrate a population of possibly bound
HVSs \citep{brown07a, brown07b}.  HVS ejection mechanisms naturally produce a broad
spectrum of ejection velocities \citep[e.g.][]{ sesana07b}.  Simulations of HVS
ejections from the Hills mechanism suggest there should be comparable numbers of
unbound and bound HVSs with $v_{rf}>+275$ km s$^{-1}$ in our survey volume
\citep{bromley06}.  We find 14 unbound HVSs and 12 possibly bound HVSs with
$v_{rf}>+275$ km s$^{-1}$, in good agreement with model predictions.

\plottwo{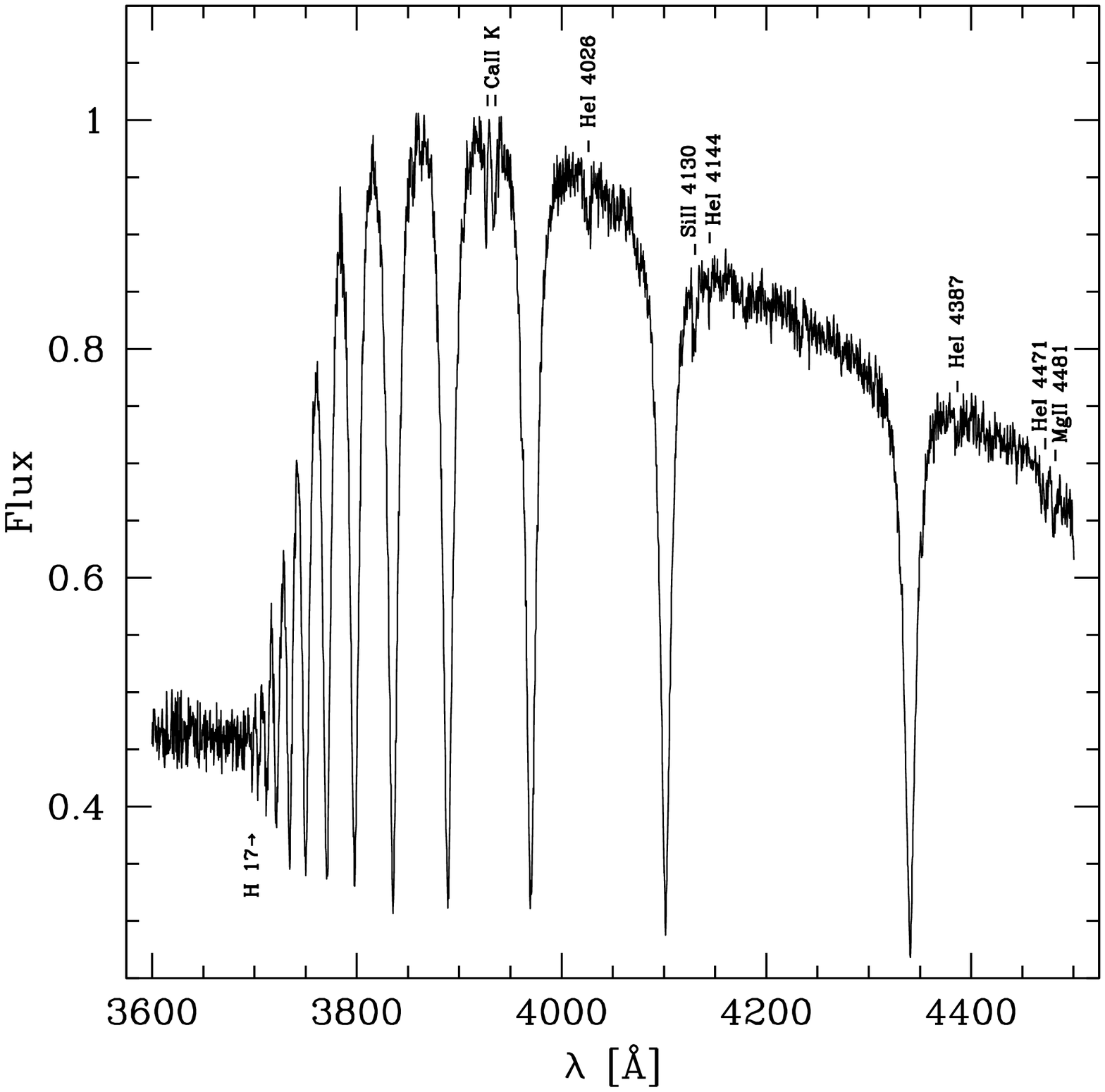}{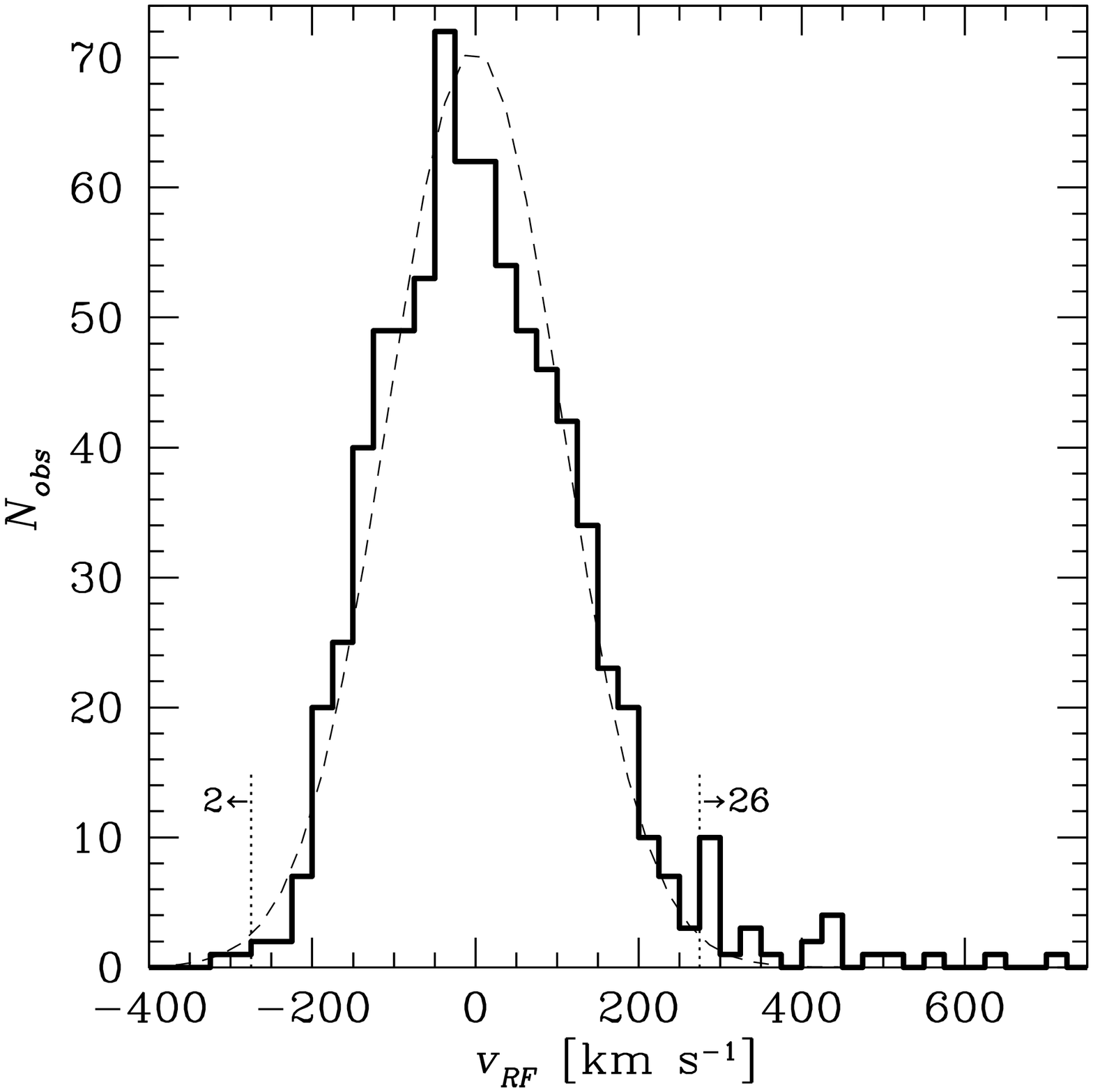}
 \figcaption{ \label{fig:spectrum} \footnotesize
	LEFT: Summed HVS spectrum, created from the weighted average of our
observations of HVS1 and HVS4 - HVS10, shifted to rest frame.  The spectral type is
that of a B8 - B9 star.  The wavelength difference between the pair of Ca {\sc ii} K
lines, one from the HVSs and one from the local interstellar medium that appears in
the spectra of HVS8 and HVS9, visibly indicates the large space motion of the HVSs,
$\Delta \lambda / \lambda \sim$ 550 km s$^{-1}$.}
 \figcaption{ \label{fig:hist} \footnotesize
	RIGHT: Observed radial velocity distribution, corrected to the Galactic
rest-frame $v_{rf}$, for the 759 stars in our HVS survey.  The best-fit Gaussian
({\it dashed line}) has dispersion $106\pm5$ km s$^{-1}$, excluding the 26 stars
with $v_{rf}>+275$ km s$^{-1}$. }

\subsection{The Nature of HVSs}

	The significant absence of stars falling back onto the Galaxy around $-300$
km s$^{-1}$ demonstrates that HVSs are short-lived \citep{brown07b, kollmeier07,
yu07}.  If bound HVSs had main sequence lifetimes greater than $\sim$1 Gyr, we would
see them falling back onto the Galaxy, contrary to the observations.  Given the
color-selection of the HVS survey, the B-type HVSs must be 3-4 M$_{\odot}$ main
sequence stars.

	Follow-up observations have confirmed that four of the HVSs are main
sequence stars:  HVS1 is a slowly pulsating B variable \citep{fuentes06}, HVS3 is a
9 M$_{\sun}$ B star \citep{bonanos08, przybilla08}, HVS7 is a 3.7 M$_{\sun}$ Bp star
\citep{przybilla08b}, and HVS8 is a rapidly rotating B star \citep{lopezmorales08}.

	The identification of HVSs as main sequence stars is in stark contrast to
the halo stars in our survey, which are, presumably, evolved 0.6-1 M$_{\sun}$ stars
on the blue horizontal branch (BHB).  BHB stars among the HVSs would be exciting,
however, because unbound BHB stars would allow us to probe the low-mass regime of
HVSs.


	Interestingly, HVS12 was previously classified as a BHB star in the BHB
samples of \citet{sirko04a} and \citet{xue08}.  The existence of $1\pm1$ BHB stars
among our 14 unbound HVSs is consistent with predictions from Galactic center
ejection models \citep{kenyon08}.
	A single HVS among the 1170 \citet{sirko04a} BHB stars and the 10224
\citet{xue08} BHB candidates also shows the immense dilution due to stars in the
Galactic halo.  Our HVS survey works because we target stars that are bluer and/or
fainter than the bulk of halo BHB stars.

\subsection{A Possible HVS from the Large Magellanic Cloud}

	HVS3, the unbound star very near the LMC on the sky, has received
considerable attention.  HVS3 is a 9 M$_{\odot}$ B star of half-solar abundance, a
good match to the abundance of the LMC \citep{bonanos08, przybilla08}.  Stellar
abundance may not be conclusive evidence of origin, however.  A- and B-type stars
exhibit 0.5 - 1 dex scatter in elemental abundances within a single cluster, due to
gravitational settling and radiative levitation in the atmospheres of the stars
\citep{varenne99, monier05, fossati07, gebran08a, gebran08b}.
	Yet because the 18 Myr lifetime of HVS3 is significantly shorter than its
travel time from the Galactic Center, HVS3 may be the first evidence for a MBH in
the LMC \citep{edelmann05}.

	An LMC origin requires that HVS3 was ejected from the galaxy at $\sim1000$
km s$^{-1}$ \citep{przybilla08}, a velocity that can possibly come from three-body
interactions with an intermediate mass black hole in a massive star cluster
\citep{gualandris07, gvaramadze08}.  \citet{perets08b} shows that the ejection rate
of 9 M$_{\sun}$ stars, however, is four orders of magnitude too small for this
explanation to be plausible.  The alternative explanation is that HVS3 is a blue
straggler, ejected by the Milky Way's MBH.  Theorists argue that a single MBH or a
binary MBH can eject a compact binary star as an HVS \citep{lu07, perets08b}; the
subsequent evolution of such a compact binary can result in mass-transfer and/or a
merger that can possibly explain HVS3 \citep{perets08b}.  Proper motion
measurements, underway now with the {\it Hubble Space Telescope}, will determine
HVS3's origin.

\section{LINKS TO THE GALACTIC CENTER}

\subsection{Ejection History}

	If Sgr A$^*$ ejects a steady fountain of HVSs from the Galactic Center, then
the space density of HVSs goes as $\rho \propto R^{-2}$ \citep{brown06, kollmeier07,
kenyon08}.  The volume sampled by a magnitude-limited survey is proportional to
$R^3$.  Thus, in the simplest picture, we expect the number of HVSs to have a linear
dependence with $R$.  Remarkably, the observed HVSs show just such a linear
cumulative distribution \citep{brown07b}.

	Figure \ref{fig:travel} explores the history of stars interacting with Sgr
A$^*$ in more detail.  We calculate Galactic Center travel times (dotted lines) from
the \citet{kenyon08} model, assuming the observed radial velocities are full space
motions.  HVS travel times span 60 - 240 Myr.

\plottwo{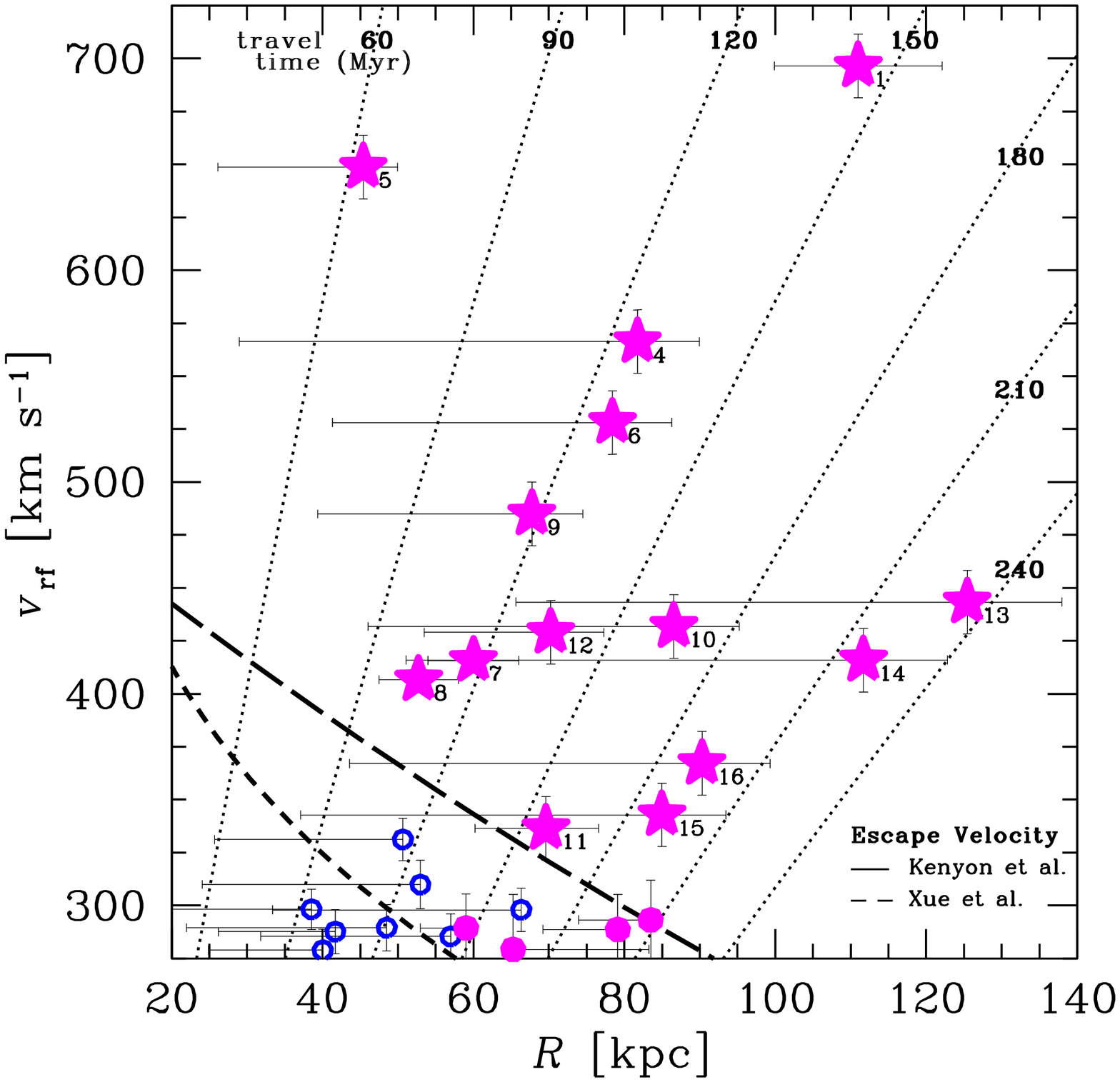}{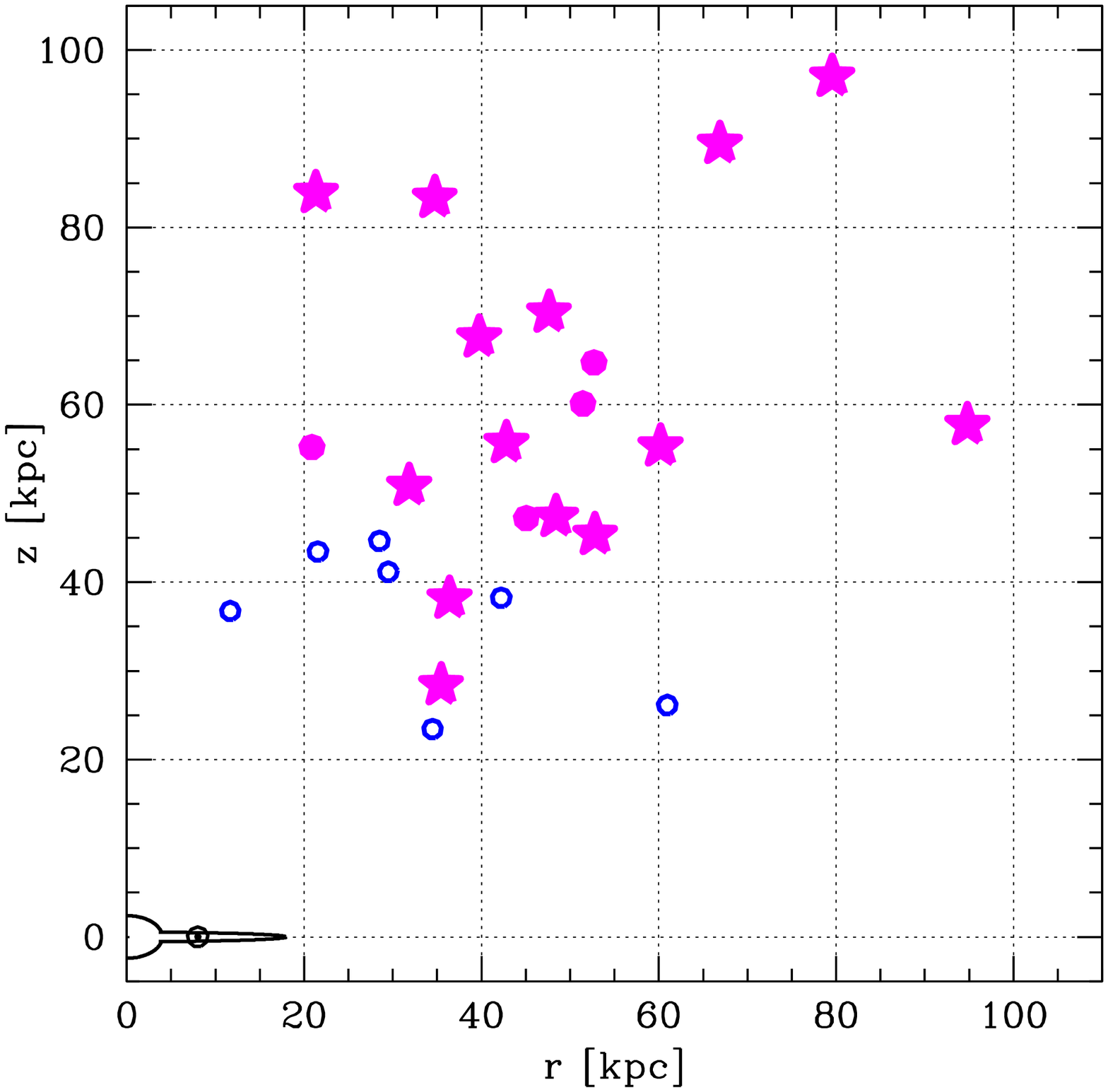}
 \figcaption{ \label{fig:travel} \footnotesize
	LEFT: Minimum rest-frame velocity vs.\ Galactocentric distance $R$ for stars
in our HVS survey \citep{brown08c}.  14 unbound HVSs ({\it solid stars}) have
velocities and distances exceeding the \citet{kenyon08} escape velocity model ({\it
long dashed line}).  4 possible HVSs ({\it dots})  are in excess of the
\citet{xue08} escape velocity model ({\it dashed line}).  8 possibly bound HVSs
({\it open circles}) are also indicated.  Errorbars show the span of physically
possible distance if the HVSs are BHB stars.  Isochrones of travel time from the
Galactic center ({\it dotted lines}) are calculated using the potential model of
\citet{kenyon08}, assuming the observed minimum rest frame velocity $v_{rf}$ is the
full space motion of the stars.  }
 \figcaption{ \label{fig:gal} \footnotesize
	RIGHT: Location of our 14 HVSs ({\it stars}), the 4 possible HVSs ({\it
dots}), and 8 possibly bound HVSs with $v_{rf}>275$ km s$^{-1}$ ({\it open
circles}). $z$ is the distance above the Galactic plane and $r$ is the distance
along the Galactic plane, such that $R=(r^2 + z^2)^{-0.5}$.  For reference, we
sketch the Milky Way and the Sun at $(r,z)=(8,0)$ kpc. }

	Statistically, the set of HVSs favor a continuous ejection process.  In
other words, there is no evidence that a massive star cluster or IMBH has
in-spiraled into the Galactic Center in the past couple hundred Myr, or at least one
that produced a coherent burst of HVSs.
	Given the small number statistics, we cannot rule out that the 5 HVSs near
$\sim$120 Myr, for example, come from a single ejection event.  However, the 5 HVSs
near $\sim$120 Myr exhibit lower velocities at shorter travel times, in the opposite
sense expected for ejections from an in-spiraling IMBH.

\subsection{Stellar Population}

	Observed HVSs exit the Galaxy in $\sim$100 Myr and so they reflect the
present stellar population near Sgr A$^*$.
	In principle, we can constrain the parent mass function of HVSs by combining
predictions of HVS rates with the observations.
	An HVS moving 500 km s$^{-1}$ travels 100 kpc in 200 Myr, thus the
\citet{yu03} rate implies $\sim$2000 HVSs of all types to a depth of 100 kpc.
	Our magnitude-limited survey reaches the same depth over 1/6 of the sky,
from which we infer $96\pm20$ unbound 3-4 M$_{\sun}$ HVSs within 100 kpc
\citep{brown07b}.  A Salpeter mass function \citep{salpeter55}, integrated over the
mass range 0.2-100 M$_{\sun}$ and normalized to 2000 stars, contains $\sim$20 stars
between 3 and 4 M$_{\sun}$.

	\citet{perets07} argues that massive perturbers, such as giant molecular
clouds, scatter stars into Sgr A$^*$'s ``loss cone'' much more efficiently than
2-body scattering.  As a result, the Galactic Center may eject HVSs at a
$\sim$10$\times$ greater rate than predicted by \citet{yu03}; in this case, the
expected number of 3-4 M$_{\sun}$ HVSs in a Salpeter mass function increases to
$\sim$200, in better agreement with observations.

	The likelihood of finding low-mass HVSs depends sensitively on the mass 
function.  
	If Sgr A$^*$ ejects stars with the present day mass function of the bulge,
there is 1 unbound F-type HVS with $g'<21.5$ per 50 deg$^2$ of sky
\citep{kollmeier07}.  
	In the Galactic Center, there is some indication that the stellar mass
function is top heavy \citep{maness07}.  A Salpeter mass function predicts an
order-of-magnitude lower density of F-type HVSs, 1 per $\sim$500 deg$^2$
\citep{brown07b}.  The ratio of high- to low-mass HVSs thus provides a sensitive
measure of the stellar mass function near Sgr A$^*$.

\subsection{Binary Fraction}

	HVSs are also linked to the properties of binaries.  In the disk, nearly all
O and B stars are in binaries, and a third of such binaries are equal-mass twins
\citep{kobulnicky07}.  It would be very interesting to know the multiplicity of
stars in the Galactic Center, particularly the S-stars orbiting Sgr A$^*$.


	If HVSs are disrupted binaries, the former companions of HVSs are left on
highly eccentric orbits around Sgr A$^*$ \citep{ginsburg06}.  The orbital properties
of the S-stars are thus linked to HVSs leaving the Galaxy today.  Stars with
main sequence lifetimes $\gtrsim$200 Myr were present when the known HVSs were
ejected from the Galactic Center.

\section{CONCLUSION}

	HVSs are fascinating because their properties are linked to Sgr A$^*$ and
the stellar environment of the Galactic Center.  
	A statistical sample of HVSs can address: 1) the nature of the MBH ejection
mechanism, 2) the in-fall history of stars onto Sgr A$^*$, 3) the types of stars
orbiting Sgr A$^*$, and 4) a unique measurement the shape of the Galaxy's dark
matter potential.  The challenge to observers is to find new HVSs and strengthen the
connection between HVSs and the Galactic Center.

\acknowledgements
	This invited review article is published in the November 2008
``Galactic Center Newsletter'' http://www.aoc.nrao.edu/$\sim$gcnews/
	I thank Margaret Geller and Scott Kenyon for their important
contributions to the HVS program.  I wish to acknowledge the contributions of
Ben Bromley and Michael Kurtz, and the financial support of the Smithsonian
Institution.

\clearpage


\begin{thebibliography}{69}
\expandafter\ifx\csname natexlab\endcsname\relax\def\natexlab#1{#1}\fi
\parskip -7pt  \scriptsize

\bibitem[{{Arzoumanian} {et~al.}(2002){Arzoumanian}, {Chernoff}, \&
  {Cordes}}]{arzoumanian02}
{Arzoumanian}, Z., {Chernoff}, D.~F., \& {Cordes}, J.~M. 2002, \apj, 568, 289

\bibitem[{{Baumgardt} {et~al.}(2006){Baumgardt}, {Gualandris}, \& {Portegies
  Zwart}}]{baumgardt06}
{Baumgardt}, H., {Gualandris}, A., \& {Portegies Zwart}, S. 2006, \mnras, 372,
  174

\bibitem[{{Blaauw}(1961)}]{blaauw61}
{Blaauw}, A. 1961, \bain, 15, 265

\bibitem[{{Bonanos} {et~al.}(2008){Bonanos}, {L{\'o}pez-Morales}, {Hunter}, \&
  {Ryans}}]{bonanos08}
{Bonanos}, A.~Z., {L{\'o}pez-Morales}, M., {Hunter}, I., \& {Ryans}, R.~S.~I.
  2008, \apjl, 675, L77

\bibitem[{{Bromley} {et~al.}(2006){Bromley}, {Kenyon}, {Geller}, {Barcikowski},
  {Brown}, \& {Kurtz}}]{bromley06}
{Bromley}, B.~C., {Kenyon}, S.~J., {Geller}, M.~J., {Barcikowski}, E., {Brown},
  W.~R., \& {Kurtz}, M.~J. 2006, \apj, 653, 1194

\bibitem[{{Brown} {et~al.}(2008){Brown}, {Geller}, \& {Kenyon}}]{brown08c}
{Brown}, W.~R., {Geller}, M.~J., \& {Kenyon}, S.~J. 2008, \apj, accepted

\bibitem[{{Brown} {et~al.}(2005){Brown}, {Geller}, {Kenyon}, \&
  {Kurtz}}]{brown05}
{Brown}, W.~R., {Geller}, M.~J., {Kenyon}, S.~J., \& {Kurtz}, M.~J. 2005,
  \apjl, 622, L33

\bibitem[{{Brown} {et~al.}(2006{\natexlab{a}}){Brown}, {Geller}, {Kenyon}, \&
  {Kurtz}}]{brown06}
---. 2006{\natexlab{a}}, \apjl, 640, L35

\bibitem[{{Brown} {et~al.}(2006{\natexlab{b}}){Brown}, {Geller}, {Kenyon}, \&
  {Kurtz}}]{brown06b}
---. 2006{\natexlab{b}}, \apj, 647, 303

\bibitem[{{Brown} {et~al.}(2007{\natexlab{a}}){Brown}, {Geller}, {Kenyon},
  {Kurtz}, \& {Bromley}}]{brown07a}
{Brown}, W.~R., {Geller}, M.~J., {Kenyon}, S.~J., {Kurtz}, M.~J., \& {Bromley},
  B.~C. 2007{\natexlab{a}}, \apj, 660, 311

\bibitem[{{Brown} {et~al.}(2007{\natexlab{b}}){Brown}, {Geller}, {Kenyon},
  {Kurtz}, \& {Bromley}}]{brown07b}
---. 2007{\natexlab{b}}, \apj, 671, 1708

\bibitem[{{Davies} {et~al.}(2002){Davies}, {King}, \& {Ritter}}]{davies02}
{Davies}, M.~B., {King}, A., \& {Ritter}, H. 2002, \mnras, 333, 463

\bibitem[{{Edelmann} {et~al.}(2005){Edelmann}, {Napiwotzki}, {Heber},
  {Christlieb}, \& {Reimers}}]{edelmann05}
{Edelmann}, H., {Napiwotzki}, R., {Heber}, U., {Christlieb}, N., \& {Reimers},
  D. 2005, \apjl, 634, L181

\bibitem[{{Eisenhauer} {et~al.}(2005)}]{eisenhauer05}
{Eisenhauer}, F. {et~al.} 2005, \apj, 628, 246

\bibitem[{{Fossati} {et~al.}(2007)}]{fossati07}
{Fossati}, L. {et~al.} 2007, \aap, 476, 911

\bibitem[{{Fuentes} {et~al.}(2006){Fuentes}, {Stanek}, {Gaudi}, {McLeod},
  {Bogdanov}, {Hartman}, {Hickox}, \& {Holman}}]{fuentes06}
{Fuentes}, C.~I., {Stanek}, K.~Z., {Gaudi}, B.~S., {McLeod}, B.~A., {Bogdanov},
  S., {Hartman}, J.~D., {Hickox}, R.~C., \& {Holman}, M.~J. 2006, \apjl, 636,
  L37

\bibitem[{{Gebran} \& {Monier}(2008)}]{gebran08b}
{Gebran}, M. \& {Monier}, R. 2008, \aap, 483, 567

\bibitem[{{Gebran} {et~al.}(2008){Gebran}, {Monier}, \& {Richard}}]{gebran08a}
{Gebran}, M., {Monier}, R., \& {Richard}, O. 2008, \aap, 479, 189

\bibitem[{{Ghez} {et~al.}(2005){Ghez}, {Salim}, {Hornstein}, {Tanner}, {Lu},
  {Morris}, {Becklin}, \& {Duchene}}]{ghez05}
{Ghez}, A.~M., {Salim}, S., {Hornstein}, S.~D., {Tanner}, A., {Lu}, J.~R.,
  {Morris}, M., {Becklin}, E.~E., \& {Duchene}, G. 2005, \apj, 620, 744

\bibitem[{{Ghez} {et~al.}(2003)}]{ghez03}
{Ghez}, A.~M. {et~al.} 2003, \apjl, 586, L127

\bibitem[{{Ginsburg} \& {Loeb}(2006)}]{ginsburg06}
{Ginsburg}, I. \& {Loeb}, A. 2006, \mnras, 368, 221

\bibitem[{{Ginsburg} \& {Loeb}(2007)}]{ginsburg07}
---. 2007, \mnras, 376, 492

\bibitem[{{Gnedin} {et~al.}(2005){Gnedin}, {Gould}, {Miralda-Escud{\'e}}, \&
  {Zentner}}]{gnedin05}
{Gnedin}, O.~Y., {Gould}, A., {Miralda-Escud{\'e}}, J., \& {Zentner}, A.~R.
  2005, \apj, 634, 344

\bibitem[{{Gould} \& {Quillen}(2003)}]{gould03b}
{Gould}, A. \& {Quillen}, A.~C. 2003, \apj, 592, 935

\bibitem[{{Gualandris} \& {Portegies Zwart}(2007)}]{gualandris07}
{Gualandris}, A. \& {Portegies Zwart}, S. 2007, \mnras, 376, L29

\bibitem[{{Gualandris} {et~al.}(2005){Gualandris}, {Portegies Zwart}, \&
  {Sipior}}]{gualandris05}
{Gualandris}, A., {Portegies Zwart}, S.~P., \& {Sipior}, M.~S. 2005, \mnras,
  363, 223

\bibitem[{{Gvaramadze} {et~al.}(2008){Gvaramadze}, {Gualandris}, \& {Portegies
  Zwart}}]{gvaramadze08}
{Gvaramadze}, V.~V., {Gualandris}, A., \& {Portegies Zwart}, S. 2008, \mnras,
  385, 929

\bibitem[{{Hansen}(2007)}]{hansen07}
{Hansen}, B.~M.~S. 2007, \apjl, 671, L133

\bibitem[{{Heber} {et~al.}(2008){Heber}, {Edelmann}, {Napiwotzki}, {Altmann},
  \& {Scholz}}]{heber08}
{Heber}, U., {Edelmann}, H., {Napiwotzki}, R., {Altmann}, M., \& {Scholz},
  R.-D. 2008, \aap, 483, L21

\bibitem[{{Hills}(1988)}]{hills88}
{Hills}, J.~G. 1988, \nat, 331, 687

\bibitem[{{Hirsch} {et~al.}(2005){Hirsch}, {Heber}, {O'Toole}, \&
  {Bresolin}}]{hirsch05}
{Hirsch}, H.~A., {Heber}, U., {O'Toole}, S.~J., \& {Bresolin}, F. 2005, \aap,
  444, L61

\bibitem[{{Kenyon} {et~al.}(2008){Kenyon}, {Bromley}, {Geller}, \&
  {Brown}}]{kenyon08}
{Kenyon}, S.~J., {Bromley}, B.~C., {Geller}, M.~J., \& {Brown}, W.~R. 2008,
  \apj, 680, 312

\bibitem[{{Kobulnicky} \& {Fryer}(2007)}]{kobulnicky07}
{Kobulnicky}, H.~A. \& {Fryer}, C.~L. 2007, \apj, 670, 747

\bibitem[{{Kollmeier} \& {Gould}(2007)}]{kollmeier07}
{Kollmeier}, J.~A. \& {Gould}, A. 2007, \apj, 664, 343

\bibitem[{{Leonard}(1991)}]{leonard91}
{Leonard}, P.~J.~T. 1991, \aj, 101, 562

\bibitem[{{Leonard}(1993)}]{leonard93}
{Leonard}, P.~J.~T. 1993, in ASP Conf.\ Ser.\ 45, Luminous High-Latitude Stars,
  ed. D.~Sasselov, 360

\bibitem[{{Leonard} \& {Duncan}(1988)}]{leonard88}
{Leonard}, P.~J.~T. \& {Duncan}, M.~J. 1988, \aj, 96, 222

\bibitem[{{Leonard} \& {Duncan}(1990)}]{leonard90}
---. 1990, \aj, 99, 608

\bibitem[{{Levin}(2006)}]{levin06}
{Levin}, Y. 2006, \apj, 653, 1203

\bibitem[{{L{\"o}ckmann} \& {Baumgardt}(2008)}]{lockmann08}
{L{\"o}ckmann}, U. \& {Baumgardt}, H. 2008, \mnras, 384, 323

\bibitem[{{L{\'o}pez-Morales} \& {Bonanos}(2008)}]{lopezmorales08}
{L{\'o}pez-Morales}, M. \& {Bonanos}, A.~Z. 2008, \apjl, 685, L47

\bibitem[{{Lu} {et~al.}(2007){Lu}, {Yu}, \& {Lin}}]{lu07}
{Lu}, Y., {Yu}, Q., \& {Lin}, D.~N.~C. 2007, \apjl, 666, L89

\bibitem[{{Maness} {et~al.}(2007)}]{maness07}
{Maness}, H. {et~al.} 2007, \apj, 669, 1024

\bibitem[{{Martins} {et~al.}(2008){Martins}, {Gillessen}, {Eisenhauer},
  {Genzel}, {Ott}, \& {Trippe}}]{martins08}
{Martins}, F., {Gillessen}, S., {Eisenhauer}, F., {Genzel}, R., {Ott}, T., \&
  {Trippe}, S. 2008, \apjl, 672, L119

\bibitem[{{Merritt}(2006)}]{merritt06}
{Merritt}, D. 2006, \apj, 648, 976

\bibitem[{{Monier}(2005)}]{monier05}
{Monier}, R. 2005, \aap, 442, 563

\bibitem[{{O'Leary} \& {Loeb}(2008)}]{oleary08}
{O'Leary}, R.~M. \& {Loeb}, A. 2008, \mnras, 383, 86

\bibitem[{{Perets}(2008)}]{perets08b}
{Perets}, H.~B. 2008, ArXiv:0802.1004

\bibitem[{{Perets} {et~al.}(2007){Perets}, {Hopman}, \& {Alexander}}]{perets07}
{Perets}, H.~B., {Hopman}, C., \& {Alexander}, T. 2007, \apj, 656, 709

\bibitem[{{Portegies Zwart}(2000)}]{portegies00}
{Portegies Zwart}, S.~F. 2000, \apj, 544, 437

\bibitem[{{Poveda} {et~al.}(1967){Poveda}, {Ruiz}, \& {Allen}}]{poveda67}
{Poveda}, A., {Ruiz}, J., \& {Allen}, C. 1967, Bol.\ Obs\ Tonantzintla
  Tacubaya, 4, 860

\bibitem[{{Przybilla} {et~al.}(2008{\natexlab{a}}){Przybilla}, {Nieva},
  {Heber}, \& {Butler}}]{przybilla08c}
{Przybilla}, N., {Nieva}, M.~F., {Heber}, U., \& {Butler}, K.
  2008{\natexlab{a}}, \apjl, 684, L103

\bibitem[{{Przybilla} {et~al.}(2008{\natexlab{b}}){Przybilla}, {Nieva},
  {Heber}, {Firnstein}, {Butler}, {Napiwotzki}, \& {Edelmann}}]{przybilla08}
{Przybilla}, N., {Nieva}, M.~F., {Heber}, U., {Firnstein}, M., {Butler}, K.,
  {Napiwotzki}, R., \& {Edelmann}, H. 2008{\natexlab{b}}, \aap, 480, L37

\bibitem[{{Przybilla} {et~al.}(2008{\natexlab{c}}){Przybilla}, {Nieva},
  {Tillich}, {Heber}, {Butler}, \& {Brown}}]{przybilla08b}
{Przybilla}, N., {Nieva}, M.~F., {Tillich}, A., {Heber}, U., {Butler}, K., \&
  {Brown}, W.~R. 2008{\natexlab{c}}, \aap, 488, L51

\bibitem[{{Reid} \& {Brunthaler}(2004)}]{reid04}
{Reid}, M.~J. \& {Brunthaler}, A. 2004, \apj, 616, 872

\bibitem[{{Salpeter}(1955)}]{salpeter55}
{Salpeter}, E.~E. 1955, \apj, 121, 161

\bibitem[{{Sch{\" o}del} {et~al.}(2003){Sch{\" o}del}, {Ott}, {Genzel},
  {Eckart}, {Mouawad}, \& {Alexander}}]{schodel03}
{Sch{\" o}del}, R., {Ott}, T., {Genzel}, R., {Eckart}, A., {Mouawad}, N., \&
  {Alexander}, T. 2003, \apj, 596, 1015

\bibitem[{{Sesana} {et~al.}(2006){Sesana}, {Haardt}, \& {Madau}}]{sesana06}
{Sesana}, A., {Haardt}, F., \& {Madau}, P. 2006, \apj, 651, 392

\bibitem[{{Sesana} {et~al.}(2007{\natexlab{a}}){Sesana}, {Haardt}, \&
  {Madau}}]{sesana07}
---. 2007{\natexlab{a}}, \apj, 660, 546

\bibitem[{{Sesana} {et~al.}(2007{\natexlab{b}}){Sesana}, {Haardt}, \&
  {Madau}}]{sesana07b}
---. 2007{\natexlab{b}}, \mnras, 379, L45

\bibitem[{{Sherwin} {et~al.}(2008){Sherwin}, {Loeb}, \& {O'Leary}}]{sherwin08}
{Sherwin}, B.~D., {Loeb}, A., \& {O'Leary}, R.~M. 2008, \mnras, 386, 1179

\bibitem[{{Sirko} {et~al.}(2004)}]{sirko04a}
{Sirko}, E. {et~al.} 2004, \aj, 127, 899

\bibitem[{{Tauris} \& {Takens}(1998)}]{tauris98}
{Tauris}, T.~M. \& {Takens}, R.~J. 1998, \aap, 330, 1047

\bibitem[{{Vanbeveren} {et~al.}(1998){Vanbeveren}, {De Loore}, \& {Van
  Rensbergen}}]{vanbeveren98}
{Vanbeveren}, D., {De Loore}, C., \& {Van Rensbergen}, W. 1998, \aapr, 9, 63

\bibitem[{{Varenne} \& {Monier}(1999)}]{varenne99}
{Varenne}, O. \& {Monier}, R. 1999, \aap, 351, 247

\bibitem[{{Wu} {et~al.}(2008){Wu}, {Famaey}, {Gentile}, {Perets}, \&
  {Zhao}}]{wu08}
{Wu}, X., {Famaey}, B., {Gentile}, G., {Perets}, H., \& {Zhao}, H. 2008,
  \mnras, 386, 2199

\bibitem[{{Xue} {et~al.}(2008)}]{xue08}
{Xue}, X. {et~al.} 2008, \apj, 684, 1143

\bibitem[{{Yu} \& {Madau}(2007)}]{yu07}
{Yu}, Q. \& {Madau}, P. 2007, \mnras, 379, 1293

\bibitem[{{Yu} \& {Tremaine}(2003)}]{yu03}
{Yu}, Q. \& {Tremaine}, S. 2003, \apj, 599, 1129

\end{thebibliography}
\end{document}